\documentclass[11pt,a4paper]{emulateapj}

\usepackage{psfrag}
\usepackage{psfig}
\usepackage{pspicture}
\usepackage{graphicx}

\usepackage{amsmath}

\def\apj{ApJ}
\def\apjl{ApJL}
\def\mnras{MNRAS}

\def\aap{A\&A}
\def\aj{AJ}
\def\apjs{ApJS}

\def\nat{Nature}

\def\gs{\mathrel{\raise0.35ex\hbox{$\scriptstyle >$}\kern-0.6em\lower0.40ex\hbox{{$\scriptstyle \sim$}}}} 
\def\ls{\mathrel{\raise0.35ex\hbox{$\scriptstyle <$}\kern-0.6em\lower0.40ex\hbox{{$\scriptstyle \sim$}}}}

\def\Wm2{\,\hbox{W}\,\hbox{m}^{-2}} 
\def\gsim{\mathrel{\raise0.35ex\hbox{$\scriptstyle >$}\kern-0.6em\lower0.40ex\hbox{{$\scriptstyle \sim$}}}} 
\def\lsim{\mathrel{\raise0.35ex\hbox{$\scriptstyle <$}\kern-0.6em\lower0.40ex\hbox{{$\scriptstyle \sim$}}}} 
\def\ltsima{$\; \buildrel < \over \sim \;$} 
\def\simlt{\lower.5ex\hbox{\ltsima}} 
\def\gtsima{$\; \buildrel > \over \sim \;$} 
\def\simgt{\lower.5ex\hbox{\gtsima}}

\lefthead{Sobral et al.}  \righthead{Dynamics of \lowercase{$z$}\,=\,0.8 H$\alpha$: KMOS/CF-HiZELS}

\begin{document}

\title {The dynamics of \lowercase{$z$}\,=\,0.8 H$\alpha$-selected star-forming galaxies from KMOS/CF-HiZELS}

\author{
D.\ Sobral,\altaffilmark{1}
A.\,M.\ Swinbank,\altaffilmark{2}
J.\,P.\ Stott,\altaffilmark{2}
J.\,J.\,A.\ Matthee,\altaffilmark{1}
R.\,G.\ Bower,\altaffilmark{2}
Ian Smail,\altaffilmark{2} \\
P.\,N.\ Best,\altaffilmark{3}
J.\,E.\ Geach,\altaffilmark{4} 
and R.\,M.\ Sharples\altaffilmark{2}}
\setcounter{footnote}{0}
\altaffiltext{1}{Leiden Observatory, Leiden University, PO Box 9513, 2300 RA Leiden, The Netherlands; email: sobral@strw.leidenuniv.nl}
\altaffiltext{2}{Institute for Computational Cosmology, Department of Physics, Durham University, South Road, Durham DH1 3LE, UK}
\altaffiltext{3}{SUPA, Institute for Astronomy, Royal Observatory of Edinburgh, Blackford Hill, Edinburgh EH9 3HJ, UK}
\altaffiltext{4}{Centre for Astrophysics Research, Science \& Technology Research Institute, University of Hertfordshire, Hatfield AL10 9AB, UK}

\begin{abstract}
We present the spatially resolved H$\alpha$ dynamics of 16 star-forming galaxies at $z\sim $\,0.81 using the new KMOS multi-object integral field spectrograph on the ESO Very Large Telescope. These galaxies, selected using 1.18\,$\mu$m narrow-band imaging from the 10\,deg$^2$ CFHT-HiZELS survey of the SA\,22\,hr field, are found in a $\sim4$\,Mpc over-density of H$\alpha$ emitters and likely reside in a group/intermediate environment, but not a cluster. We confirm and identify a rich group of star-forming galaxies at $z= $\,0.813\,$\pm$\,0.003, with 13 galaxies within 1000\,km\,s$^{-1}$ of each other, and seven within a diameter of 3\,Mpc. All our galaxies are ``typical'' star-forming galaxies at their redshift, 0.8\,$\pm0.4$\,SFR$^*_{z=0.8}$, spanning a range of specific star formation rates (sSFRs) of 0.2--1.1\,Gyr$^{-1}$ and have a median metallicity very close to solar of 12\,+\,log(O/H)\,=\,8.62\,$\pm$\,0.06. We measure the spatially resolved H$\alpha$ dynamics of the galaxies in our sample and show that 13 out of 16 galaxies can be described by rotating disks and use the data to derive inclination corrected rotation speeds of 50--275\,km\,s$^{-1}$. The fraction of disks within our sample is 75\%\,$\pm$\,8\%, consistent with previous results based on \emph{HST} morphologies of H$\alpha$-selected galaxies at $z\sim $\,1 and confirming that disks dominate the star formation rate density at $z\sim $\,1. Our H$\alpha$ galaxies are well fitted by the $z\sim1-2$ Tully-Fisher (TF) relation, confirming the evolution seen in the zero-point. Apart from having, on average, higher stellar masses and lower sSFRs, our group galaxies at $z=0.81$ present the same mass-metallicity and TF relation as $z\sim1$ field galaxies and are all disk galaxies.

\end{abstract}


\section{Introduction}

The properties of star forming galaxies have changed dramatically in the 7\,Gyr between $z$\,=\,1 and the present day \citep[e.g.\ ][]{Madau96,Sobral09}.  In particular, the comoving star formation rate density of the Universe has dropped by an order of magnitude over this time \citep{Rodighiero11,Karim11,Gilbank11,Sobral13}. The decline affects the star forming population at all masses \citep{Sobral13b}, and is much more rapid than predicted by galaxy formation models \citep{Cirasuolo10,Bower12}.

Two theories have been suggested to explain this rapid decline: {(1)} the rate of mergers and tidal interactions may be higher at $z\sim $\,1\,--\,2, driving quiescent disks into bursts of star formation \citep[e.g.\ ][]{Conselice09}; {(2)} gas accretion rates are much higher at $z$\,=\,1\,--\,2, leading to higher gas densities and consequently higher star formation rates \citep[e.g.\ ][]{Dekel09}.  Whichever process dominates the gas accretion onto galaxies at high redshift, it appears that the higher rate of halo growth, together with lower specific angular momentum for fixed circular velocity \citep[][]{Dutton11} results in gas disks that are intrinsically more unstable -- unless counter balanced by high star formation rates and turbulence \citep[e.g.\ ][]{Hopkins12b,Swinbank12b,Livermore12a}.  

Significant effort has been invested to measure the velocity motions of the gas within star-forming galaxies at $z\sim $\,1--2 in order to test competing models for galaxy growth \citep[e.g.\ see the recent review by][]{Glazebrook13}. In particular, it appears that the majority of star forming systems at $z\sim $\,1\,--\,2 are supported by highly turbulent, rotationally supported disks with star formation that is significantly clumpier than comparably luminous galaxies at $z\sim $\,0 \citep{Elmegreen09,Schreiber09,Genzel10,Wisnioski11,Swinbank12b}.

%
%
\begin{figure*}
  \centering
  \includegraphics[width=18cm]{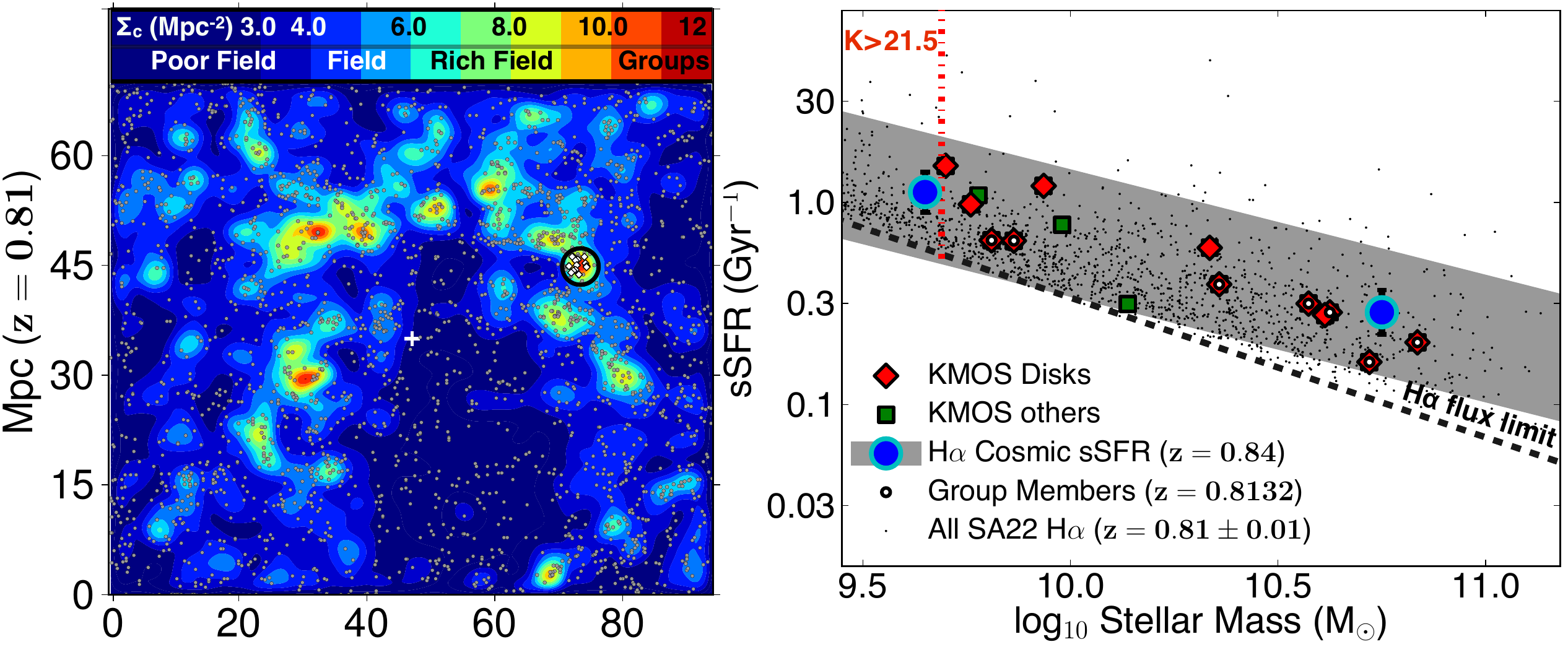}
\caption{ {\it Left:} The on-sky distribution (center of the field: 22:15:34 $+$00:20:56) of the entire sample of $\sim3000$ H$\alpha$ emitters (black dots) when compared to (corrected) local projected densities ($\Sigma_c$; 10th nearest neighbour) at $z\sim0.8$ within the SA22 field. Local densities are based on the combination of the H$\alpha$ emitters and a robust photo-z selected sample ($\sim 15$\,k sources) at $z\sim0.8$ within the SA22 field and takes into account the likely contamination and completeness of the photo-z sample following \cite{Sobral11}. Using KMOS, we have observed the largest over-density of H$\alpha$ emitters (black circle). Based on our local density estimates, and following \cite{Sobral11}, these galaxies likely reside in a group environment, but not in a cluster. {\it Right:} The relation between specific star formation rates and stellar mass for our CFHiZELS KMOS galaxies. We also show how our KMOS galaxies compare with the entire CFHT H$\alpha$ parent sample of $\sim3000$ H$\alpha$ emitters and show the sources that are within a physical diameter of $\sim3$\,Mpc at $z=0.8132$ (Group members). Note that the group members present sSFRs systematically lower than the rest of the KMOS sources which are on the outskirts of the structure ($\sim10-100$ Mpc away) and are also more massive than the rest of the sources. We also show the relation between cosmic sSFR (the ratio between the star formation rate density, $\rho_{SFR}$, by galaxies within a mass bin and the integral of the mass function within that mass bin, $\rho_*$) and Mass from the HiZELS survey at $z=0.84$ \citep{Sobral13b}. KMOS galaxies sample both a wide range in mass and sSFR. Our sample is H$\alpha$ selected, and thus we indicate the approximate flux limit of the parent sample to illustrate the region where our sample is complete.}

\label{fig:targets}
\end{figure*}

To chart the evolution and large scale clustering of star-forming galaxies with cosmic time, we have recently conducted a large (10-square degree) narrow-band survey in SA22 using the 1.18\,$\mu$m (lowOH2) narrow-band filter on WIRcam/CFHT, mostly focused on obtaining the largest samples of H$\alpha$ star-forming galaxies at $z=0.81\pm0.01$ (CF-HiZELS). Due to the depth achieved by our observations ($\sim$0.2\,L$^*_{z=0.8}$), the majority of our selected galaxies have properties ``typical'' of galaxies which will likely evolve into $\sim$\,L$^{\star}$ (or SFR$^*$) galaxies by \emph{z}\,=\,0. This survey builds on our previous successful H$\alpha$ narrow-band imaging of $\sim1$-square degree areas in redshift slices at $z$\,=\,0.40, 0.84, 1.47 and 2.23 \citep[][]{Geach08,Sobral09,Sobral12,Sobral13} from HiZELS. 

The large HiZELS samples of H$\alpha$ emitters have yielded the first self-consistent determination of the H$\alpha$ luminosity function since $z=2.23$ and show that the bulk of its evolution is driven by a strong evolution in L$^*$. HiZELS is also making important contributions towards unveiling the nature and evolution of star-forming galaxies over the last 11\,Gyrs  \citep{Sobral10,Garn10,Sobral11,Sobral12,Geach12,Swinbank12a,Swinbank12b,Stott13a,Ibar13}.

Within the CF-HiZELS survey of the SA22 field, we have identified a significant ($\sim8\sigma$) over-density of H$\alpha$ emitters within a 3000\,Mpc$^3$ volume (co-moving). To compile resolved dynamics, measure the disk turbulence and rotation speed of galaxies within this volume, we have obtained the spatially resolved H$\alpha$ measurements with the KMOS Integral field spectrograph \citep{Sharples13} during science verification time\footnote{http://www.eso.org/sci/activities/vltsv/kmossv.html}. In this paper, we use these data to investigate the dynamical properties of the galaxies, the evolution of the luminosity and stellar mass scaling relations (through the Tully-Fisher relation), and the star formation and enrichment within their ISM.  We use a cosmology with $\Omega_{\Lambda}$\,=\,0.73, $\Omega_{m}$\,=\,0.27, and H$_{0}$\,=\,72\,km\,s$^{-1}$\,Mpc$^{-1}$. In this cosmology, at the redshift of our survey, $z$\,=\,0.81, a spatial resolution of 0.5$''$ corresponds to a physical scale of $\approx $\,4\,kpc. All quoted magnitudes are on the AB system and we use a Chabrier IMF.

\section{Sample Selection, Observations \& Data Reduction}

\subsection{CFHIZELS: A Contiguous 10$\,$deg$^2$ NB Survey}

We have used the narrow-band (NB) lowOH2 filter ($\lambda=1187\pm5$\,nm) on WIRCam\,/\,CFHT \citep{Puget2004}, to image a 10\,deg$^2$ contiguous area in the SA22 \citep{Sobral13c}. This represents by far the largest contiguous narrow-band survey for high-$z$ star-forming galaxies yet undertaken and results in the largest sample of $z\sim $\,1--2 line emitters. Indeed, the survey yields $\sim3000$ robust H$\alpha$ emitters at $z$\,=\,0.81\,$\pm$\,0.01 (see Sobral et al. 2013c for details on the spec-$z$, photo-$z$ and color-color selection). As can be seen in Fig.~\ref{fig:targets}, there appears to be a significant large-scale over-density of H$\alpha$ emitters which contains $\sim $\,300 candidate $z=0.81$ H$\alpha$ line emitters within a $\sim $\,20-arcmin field (Fig.~\ref{fig:targets}). This includes a region where the number density of H$\alpha$ emitters is $\sim10$ times higher than the general field, and thus ideally suited to KMOS.

In order to investigate the physical environment in which H$\alpha$ galaxies reside, we have computed local environmental densities based on the 10th nearest neighbour and following \cite{Sobral11}\footnote{We use a sample of 15432 galaxies at $z=0.81$, which includes all the H$\alpha$ emitters, but also photo-z selected galaxies ($0.77<photo_z<0.83$). Following the method described in \cite{Sobral11}, we also apply corrections for the contamination ($\sim60$\%) and completeness ($\sim70$\%) of the photo-z sample when compared to the H$\alpha$ redshift distribution of $z=0.81\pm0.01$.}. We show local environment densities in Fig.~\ref{fig:targets} and a comparison to the distribution of all the H$\alpha$ emitters. We find that the H$\alpha$ emitters we have observed with KMOS likely reside in a group environment (see Fig.~\ref{fig:targets}), but not a cluster \citep[c.f.][]{Sobral11}, and that H$\alpha$ emitters avoid the highest local densities in the entire field.

We use the wealth of ancillary data, including 7-band photometric coverage (from $u$ to $K$-band) to compute stellar masses for all of the H$\alpha$ emitters in the parent sample following \citet{Sobral11,Sobral13b}. Due to the lack of {\it Spitzer}\,/\,IRAC data, we find that the derived masses have errors of approximately 0.2--0.3\,dex. In order to test whether the unavailability of IRAC data leads to any systematic offset in masses (and correct for it), we take the HiZELS sample of H$\alpha$ selected star-forming galaxies at $z=0.84$ \citep[COSMOS+UDS;][]{Sobral13}, apply the same selection as our H$\alpha$ sample in SA22, and derive stellar masses with only the bands we have access to in SA22 ($ugrizJK$). We compare them with masses derived with all the bands, including IRAC, and find that apart from the individual errors/scatter increasing (confirming the errors we estimate of 0.2--0.3\,dex), there is a systematic difference of +0.075\,dex for masses derived without IRAC when compared to those with IRAC for these $z=0.8$ H$\alpha$ emitters. Once we correct for that systematic offset (mass overestimation) the masses agree very well. We also find an excellent agreement between the volume-averaged mass distribution of our SA22 sources (without IRAC, but correcting for the systematic offset) and those from HiZELS (COSMOS+UDS, with all the bands) with exactly the same selection function.

%
%
%
\begin{table*}
\begin{center}
{\scriptsize
{\centerline{\sc Table 1: Integrated Galaxy Properties}}
\begin{tabular}{lccccccccccccc}
\hline
\noalign{\smallskip}
ID        & $\alpha_{\rm J2000}$ & $\delta_{\rm J2000}$ & $z$    & $K_{\rm AB}$ & R$_{1/2K}$ & [N{\sc ii}]\,/\,H$\alpha$  & Stellar Mass & SFR                    & $i$    & $v_{\rm 80}$           & $\sigma$        & K$_{\rm TOT}$ \\
            &                    &                     &        &            &      kpc         &                            &              log(M$_{\odot}$)              & $M_{\odot}$\,yr$^{-1}$ &        & km\,s$^{-1}$          & km\,s$^{-1}$      &             \\
\hline
1709 & 22:19:31.92 & +00:36:11.57        & 0.8133 & 21.3        &  $2.1\pm0.3 $          &  0.42\,$\pm$\,0.06        & $10.7\pm0.1$                      &  8.5                    &  42   &    55.\,$\pm$\,17      &  92\,$\pm$\,9    & 0.46\,$\pm$\,0.10   \\
1713 & 22:19:21.34 & +00:36:42.70        & 0.7639 & 21.1        &  $3.9\pm0.4  $         &  ...                                     & $10.0\pm0.2$                      &  7.4                   &  60   &         ...            &  33\,$\pm$\,6    & ...                 \\ 
1721 & 22:19:24.10 & +00:37:11.16        & 0.8144 & 20.0        &  $5.1\pm0.2  $         &  0.62\,$\pm$\,0.06        & $10.8\pm0.1$                      & 13.9                    &  50   &    240\,$\pm$\,14      &  66\,$\pm$\,8    & 0.60\,$\pm$\,0.23   \\
1724 & 22:19:27.27 & +00:37:31.26        & 0.8117 & 21.4        &  $4.7\pm 0.7  $        &  0.36\,$\pm$\,0.08        &  $10.1\pm0.1 $                    &  4.3                    &  50   &         ...            &  63\,$\pm$\,8    & ...                 \\
1733 & 22:19:43.57 & +00:38:22.14        & 0.7731 & 22.2        &  $3.8\pm 0.7  $        &  0.19\,$\pm$\,0.03        &  $9.7\pm0.3 $                   &  7.6                    &  63   &    90.\,$\pm$\,15      &  86\,$\pm$\,9    & 1.36\,$\pm$\,0.50   \\ 
1739 & 22:19:42.27 & +00:38:31.57        & 0.8042 & 20.1        &  $6.0\pm 0.2  $        &  0.40\,$\pm$\,0.05        & $10.6 \pm0.2  $                   &  11.4                    &  50   &    247\,$\pm$\,15      &  53\,$\pm$\,5    & 0.46\,$\pm$\,0.12   \\
1740 & 22:19:18.60 & +00:38:43.89        & 0.8128 & 21.2        &  $5.0\pm 0.4 $         &  0.32\,$\pm$\,0.05        & $10.4 \pm0.1   $                  &  8.9                    &  42   &    217\,$\pm$\,10      &  83\,$\pm$\,10   & 0.28\,$\pm$\,0.05   \\
1745 & 22:19:29.51 & +00:38:52.07        & 0.8174 & 22.0        &  $4.1 \pm 0.5 $        &  0.16\,$\pm$\,0.02        &  $9.8  \pm0.3    $                & 5.6                    &  46   &    211\,$\pm$\,20      &  60\,$\pm$\,6    & 0.19\,$\pm$\,0.10   \\
1759 & 22:19:41.42 & +00:39:25.37        & 0.8035 & 20.3        &  $4.1 \pm 0.2 $        &  0.39\,$\pm$\,0.03        & $10.3\pm0.2    $                  &  12.9                    &  38   &    275\,$\pm$\,18      &  71\,$\pm$\,6    & 0.19\,$\pm$\,0.13   \\
1770 & 22:19:27.66 & +00:40:14.30        & 0.7731 & 21.7        &  $3.9\pm  0.5 $        &  0.05\,$\pm$\,0.01        &  $9.9 \pm0.3    $                 &  10.4                    &  42   &    144\,$\pm$\,15      &  61\,$\pm$\,7    & 0.42\,$\pm$\,0.20   \\ 
1774 & 22:19:30.59 & +00:40:31.52        & 0.8127 & 21.7        &  $3.8\pm 0.5$          &  0.19\,$\pm$\,0.03        &  $9.8\pm0.2     $                 &  4.2                    &  57   &    50.\,$\pm$\,12      &  86\,$\pm$\,9    & 0.25\,$\pm$\,0.09   \\
1787 & 22:19:39.21 & +00:41:20.80        & 0.8132 & 20.5        &  $6.5\pm  0.2$         &  0.41\,$\pm$\,0.04        & $10.6\pm0.2    $                  &  12.0                    &  37   &    255\,$\pm$\,15      &  77\,$\pm$\,9    & 0.33\,$\pm$\,0.10   \\
1789 & 22:19:23.19 & +00:41:23.83        & 0.8130 & 20.6        &  $9.5\pm  0.4$         &  0.32\,$\pm$\,0.02        & $10.6\pm0.1    $                  &  11.8                    &  34   &    253\,$\pm$\,15      &  44\,$\pm$\,6    & 0.11\,$\pm$\,0.13   \\
1790 & 22:19:24.69 & +00:41:26.09        & 0.8124 & 22.0        &  $1.7 \pm 1.7$         &  0.30\,$\pm$\,0.05        & $9.9\pm0.3      $               &  4.7                    &  40   &    30.\,$\pm$\,10      &  44\,$\pm$\,6    & 0.40\,$\pm$\,0.23   \\
1793 & 22:19:30.60 & +00:41:35.12        & 0.8161 & 21.3        &  $9.3\pm0.6$           &  0.30\,$\pm$\,0.04        & $10.2\pm0.2$                      &  7.8                    &  14   &         ...            &  75\,$\pm$\,6    & ...                 \\
1795 & 22:19:32.44 & +00:41:42.32        & 0.8095 & 21.5        &  $3.0\pm0.4$           &  0.32\,$\pm$\,0.04        &  $9.8\pm0.2$                     &  6.5                    &  75   &    53.\,$\pm$\,10      &  49\,$\pm$\,5    & 0.45\,$\pm$\,0.11   \\
\hline
\label{table:Gal_props}
\end{tabular}
}
\caption{Notes: r$_{1/2}$ is the K-band (UKIDSS DXS) half light radius and has been deconvolved for the PSF ($\approx0.78''$).  $v_{\rm 80}$ is the inclination corrected rotation speed at $r_{80}$ ($r_{80}$\,=\,2.2\,r$_{1/2}$). SFRs are derived from H$\alpha$ luminosities and corrected for dust extinction by following Garn \& Best (2010). $\sigma$ denotes the average line of sight velocity dispersion (corrected for the velocity gradient of the galaxy across the PSF). K$_{\rm Tot}$ is the total kinemetric asymmetry, $K_{\rm Tot}^2$\,=\,K$_{\rm  V}^2$\,+\,K$_{\sigma}^2$.}
\end{center}
\end{table*}

\subsection{KMOS Observations}

To measure the dynamics of these galaxies we used the unique multiplexing capability of the new KMOS spectrograph which consists of 24 integral field units (IFUs) that can be deployed across a 7.2\,arcminute field. Each IFU covers an area of 2.8$''\times2.8''$ sampled by $0.2''\times0.2''$ spatial pixels. Within the over-density (Fig.~\ref{fig:targets}), we have identified 30 H$\alpha$ emitters which lie within a 7$'$ diameter region, 20 of which are brighter than $K_{\rm AB}$\,$\sim$\,21.5 (roughly corresponding to stellar mass $M_{\star}> $\,10$^{9.75}$\,$M_{\odot}$, see Fig.~\ref{fig:targets}) and have H$\alpha$ fluxes (estimated from our NB survey) brighter than 1\,$\times$\,10$^{-16}$\,erg\,s$^{-1}$\,cm$^{-2}$, (star formation rates $>2.5$\,M$_{\odot}$\,yr$^{-1}$, assuming 1 mag of extinction). We therefore selected 20 H$\alpha$ emitters for observations during science verification time with KMOS. The galaxies in this KMOS sample have a median stellar mass of $\approx10^{10.2}$\,M$_{\odot}$\,yr$^{-1}$, a median SFR of 6\,M$_{\odot}$\,yr$^{-1}$ (after correcting for extinction following \cite{GarnBest10} -- see also \cite{Sobral12,Dominguez13,Ibar13}) and a median sSFR of 0.47\,Gyr$^{-1}$.  Fig.~\ref{fig:targets} shows how these compare with both the parent population of H$\alpha$ emitters in the full SA22 field, but also when compared to other $z$\,=\,0.84 HiZELS data \citep[][]{Sobral09,Sobral13}. Our KMOS sources are typical star-forming galaxies at their redshift (4-14\,M$_{\odot}$\,yr$^{-1}$, while the typical SFR [SFR$^*$] at $z\sim0.8$ is $\sim10$\,M$_{\odot}$\,yr$^{-1}$), and provide a range in sSFR: 0.2--1.1\,Gyr$^{-1}$.

%
%
%
\begin{figure*}
  \centering
 \includegraphics[width=7.08in]{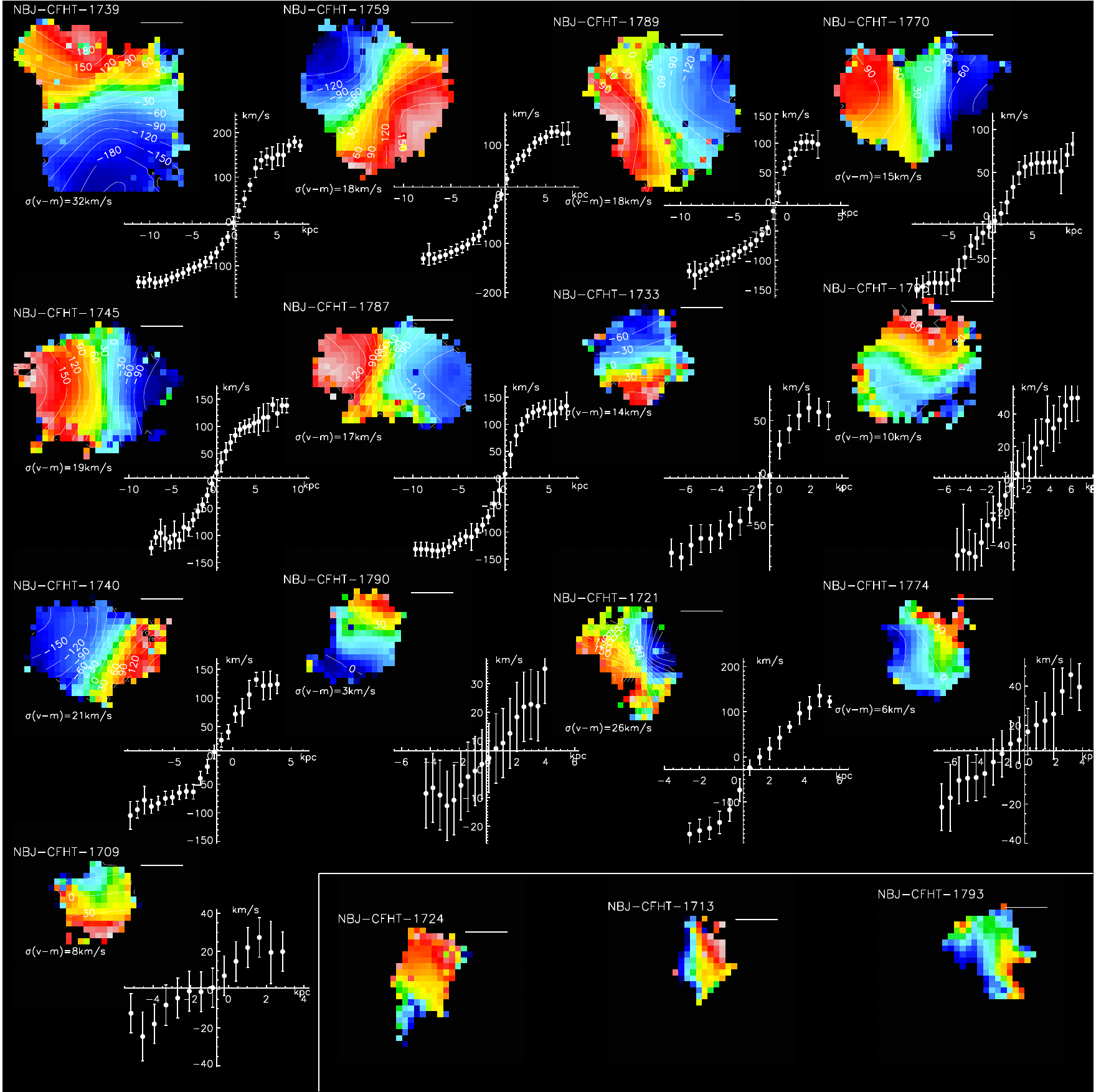}
\caption{Two dimensional velocity fields for the sixteen galaxies in our KMOS sample. The contours denote the dynamics of the best-fit two dimensional disk model.  From these velocity fields, thirteen galaxies have dynamics that resemble rotating systems, and we extract one dimensional rotation curves (shown as insets for each galaxy) based on the dynamical center and position angle from the best-fit dynamical model. In these plots, the error bars for the velocities are derived from the formal $1\sigma$ uncertainty in the velocity arising from the Gaussian profile fits to the H$\alpha$ emission. For the final three galaxies in this plot neither of the kinemetry calculation nor the disk modelling converged, and thus we do not attempt to derive rotation speeds in these three systems. We show the PSF size as a line next to each source for comparison.}
\label{fig:2dmaps}
\end{figure*}

KMOS observations were taken in 2013 June 29 and July 1.  During the observations the average $J$-band seeing was approximately 0.7$''$. We used the $YJ$-band grating to cover the H$\alpha$ emission, which at $z\sim$\,0.81 is redshifted to $\approx$\,1.187$\mu$m.  In this configuration, the spectral resolution (measured from the skylines at 1.2$\mu$m) is R\,=\,$\lambda$\,/\,$\Delta\lambda$\,=\,3430.  We also deployed three IFUs to (blank) sky positions to improve the sky-subtraction during the data reduction. Observations were carried out using an ABA sequence in which we chopped by 5$''$ to sky, and each observation was dithered by up to 0.2$''$. During the observations, three of the IFUs were disabled and so 18 galaxies were observed.

To reduce the data, we used the esorex\,/\,{\sc spark} pipeline \citep{Davies13} which extracts the slices from each IFU, flatfields and wavelength calibrates the data and forms the datacube. We reduced each AB pair separately, and improved the sky OH subtraction in each AB pair for each IFU using the data from the sky IFU from the appropriate spectrograph (using the sky-subtraction techniques described in \citealt{Davies07}). We then combined the data into the final datacube using a clipped average. The total exposure time (sky+targets) was 7.2\,ks. We note that both the effects of instrumental resolution and the spatial PSF are fully taken into account throughout the analysis and included in the error estimation.

\subsection{Galaxy Dynamics}

From the reduced data, we first collapse each datacube into a one dimensional spectrum and measure the redshift and H$\alpha$ and [N{\sc ii}] line flux (Table~1). The two faintest galaxies in our sample are only weakly detected with S/N$<$5 in H$\alpha$ so we will not use them, leaving us with a sample of 16 robustly detected galaxies.  

To measure the H$\alpha$ dynamics of each galaxy, we fit the H$\alpha$ and [N{\sc ii}]$\lambda\lambda$6548,6583 emission lines spaxel-to-spaxel using a $\chi^{2}$ minimisation procedure (and accounting for the increased noise at the positions of the sky lines). We start by trying to identify a line in a 0.4\,$\times$\,0.4$''$ region ($\sim$\,3\,kpc), and if the fit fails to detect the emission line, the region is increased to 0.6\,$\times$\,0.6$''$. We require a signal-to-noise $>$\,5 to detect the emission line. When this criterion is met we fit the H$\alpha$ and [N{\sc ii}]$\lambda\lambda$\,6548,6583 emission lines allowing the centroid, intensity and width of the Gaussian profile to find their optimum fit (the FWHM of the H$\alpha$ and [N{\sc ii}] lines are coupled in the fit).  Uncertainties in each parameter are then calculated by perturbing each parameter, one at a time, allowing the remaining parameters to find their optimum values, until $\Delta\chi^2$\,=\,1 is reached.

In Fig.~\ref{fig:2dmaps} we show the velocity fields for each of the sixteen galaxies in our final sample. All of these galaxies display velocity gradients in their dynamics, with peak-to-peak differences ranging from $\Delta v$\,=\,40--300\,km\,s$^{-1}$.  

Many of these galaxies have H$\alpha$ velocity fields which resemble rotating systems (characteristic ``spider'' patterns in the velocity fields and line of sight velocity dispersion profiles which peak near the central regions).  Therefore, we attempt to model the two dimensional velocity field to identify the dynamical center and kinematic major axis.  We follow \citet{Swinbank12a} to construct two dimensional models with an input rotation curve following an {\rm
  arctan} function [$v(r)$\,=\,$\frac{2}{\pi}$\,$v_{\rm asym}$\,${\rm arctan}(r/r_t)$], where $v_{\rm asym}$ is the asymptotic rotational velocity and r$_{\rm t}$ is the effective radius at which the rotation curve turns over \citep{Courteau97}.  Briefly, the suite of two dimensional models we fit have six free parameters ([x,y] center, position angle (PA), $r_{\rm t}$, v$_{\rm asym}$, and disk inclination) and we use a genetic algorithm \citep{Charbonneau95} to find the best model \citep[see][]{Swinbank12a}.

The best-fit kinematic maps for galaxies which can be adequately described by a rotation disk are also shown in Fig.~\ref{fig:2dmaps}. We note that all of the galaxies show small-scale deviations from the best-fit model, as indicated by the typical r.m.s, $<$\,data\,$-$\,model\,$>$\,=\,20\,$\pm$\,5\,km\,s$^{-1}$, with a range from $<$\,data\,$-$\,model\,$>$\,=\,5--30\,km\,s$^{-1}$. These offsets
could be caused by the effects of gravitational instability, or simply due to the unrelaxed dynamical state indicated by the high velocity dispersions ($\sigma$\,=\,65\,$\pm$\,6\,km\,s$^{-1}$).

Using the best fit dynamical model, we use the dynamical center and position angle of the disk and extract the one dimensional rotation curve and velocity dispersion profiles from the major kinematic axis of each galaxy and also show these in Fig.~\ref{fig:2dmaps}.  Despite the short integration time (less than 2\,hrs on source), the data clearly yield rotation curves which turn over (or flatten) for at least nine of these galaxies, clearly demonstrating the capabilities of KMOS.

While the dynamical modelling provides a useful means of identifying the major kinematic axis and dynamical center for the galaxy, another useful criterion for distinguishing between rotation and motion from disturbed kinematics is the ``kinemetry'' (which measures the asymmetry of the velocity field and line-of-sight velocity dispersion maps for each galaxy).  Kinemetry has been well calibrated and tested at low redshift \citep[e.g.\ ][]{Krajnovic07}, and also used at high redshift to determine the strength of deviations of the observed velocity and dispersion maps from an ideal rotating disk (\citealt{Shapiro08}; \citealt{Swinbank12a}; but see also \citealt{Goncalves10}).  Briefly, in this modelling, the velocity and velocity dispersion maps are described by a series of concentric ellipses of increasing semi-major axis length, as defined by the system center, position angle and inclination.  Along each ellipse, the moment map as a function of angle is extracted and decomposed into its Fourier series which have coefficients $k_n$ at each radii (see \citealt{Krajnovic07} for more details).

We measure the velocity field and velocity dispersion asymmetry for all of the galaxies in our sample, defining the velocity asymmetry (K$_{\rm V}$) and the velocity dispersion asymmetry (K$_{\sigma}$). For an ideal disk, the values of K$_{\rm V}$ and K$_{\sigma}$ will be zero. In contrast, in a merging system, strong deviations from the idealised case causes large values of K$_{\rm V}$ and K$_{\sigma}$ (which can reach K$_{\rm V}\sim$\,K$_{\sigma}\sim$\,10 for very disturbed systems).  The total asymmetry, K$_{\rm Tot}$ is $K_{\rm Tot}^2$\,=\,K$_{\rm  V}^2$\,+\,K$_{\sigma}^2$.

For the KMOS galaxies in our sample, we measure the velocity and velocity dispersion asymmetry and report their values in Table~1. NBJ-CFHT 1724, 1713 and 1793 have too few independent spatial resolution elements across the galaxy and neither the kinemetry calculation nor the disk modelling converged, so we omit the dynamical properties of these galaxies from the analysis.  Although the error bars on K$_{\rm Tot}$ are large (these errors are found by bootstrap resampling for the errors in the velocities, velocity dispersions and dynamical centers of each galaxy), the average K$_{\rm Tot}$\,=\,0.40\,$\pm$\,0.07 suggests that the majority of these galaxies are dominated by disk-like dynamics (indeed, twelve of the thirteen galaxies in our sample have K$_{\rm Tot} <$\,0.5).

%
%
\begin{figure}
 \centering
  \includegraphics[width=3.4in]{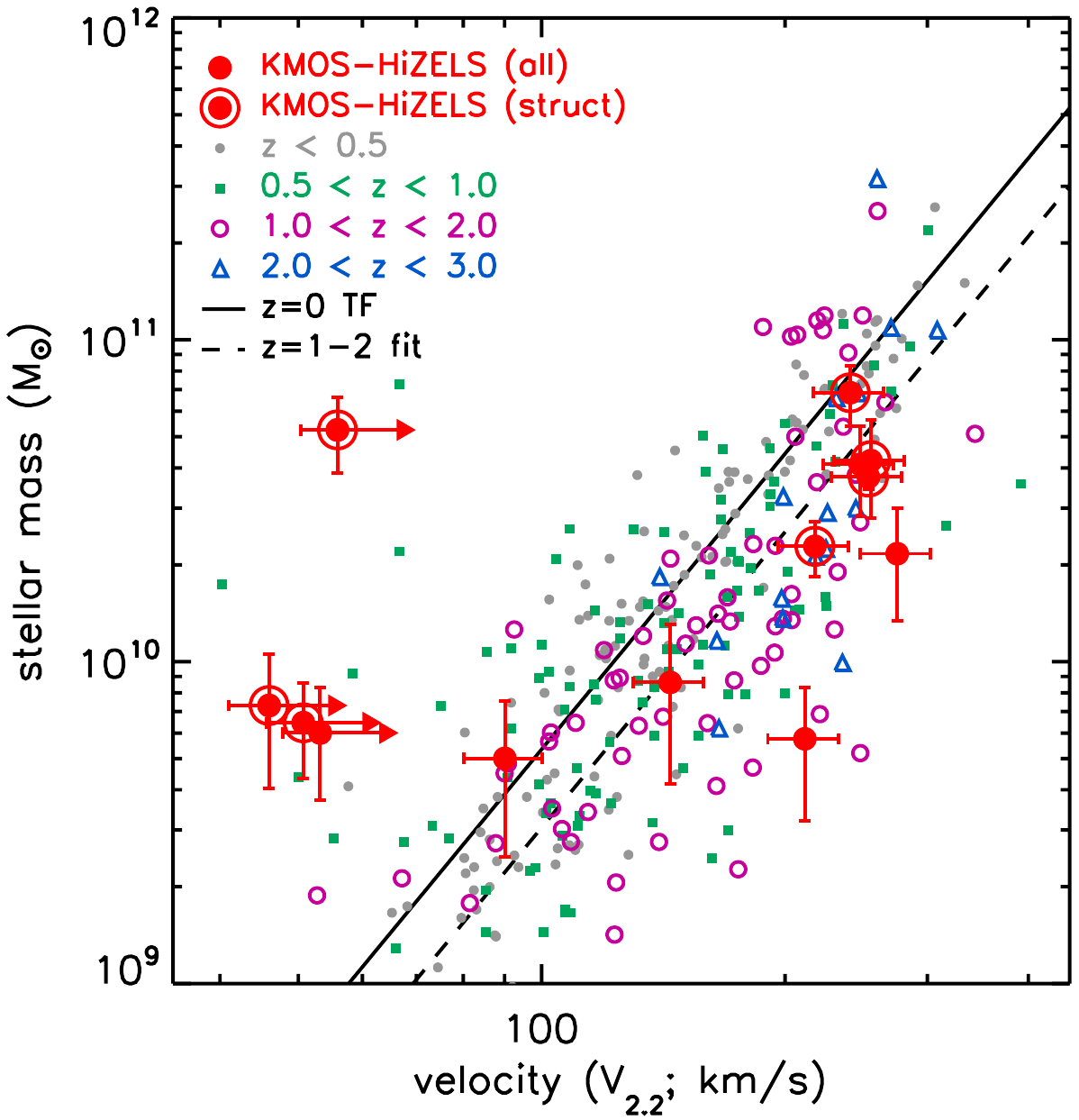}
\caption{The evolution of the stellar mass Tully-Fisher relation for all our SA\,22 KMOS sample at $z\sim $\,0.8 and for those within a 3\,Mpc diameter region (circles points). For the majority of our KMOS sources (nine) galaxies we resolve the turn over, while for the remaining four galaxies that are consistent with rotating disks but for which we do not resolve the turn over we present them as limits. We do not show the 3 galaxies for which a rotating disk is not a good fit/model. Our results are compared to a number of other low- and high- redshift surveys. The stellar masses and velocities from the literature have been estimated in a consistent way, and these values (or corrections, where necessary) are presented in \cite{Swinbank12a}. We also show the TF relation at $z=0$ and the best fit TF relation at $z=1-2$ from the compilation of star-forming galaxies from \cite{Swinbank12a}. The $z$\,=\,0 baseline for this comparison is taken from \citet{Pizagno05}, whilst the high-redshift points are from \citet{Miller11,Miller12} ($z$\,=\,0.6--1.3); \citet{Swinbank06a} ($z$\,=\,1); \citet{Swinbank12a} ($z$\,=\,1.5); \citet{Jones10} ($z$\,=\,2); \citet{Cresci09} ($z$\,=\,2) and \citet{Gnerucci11} ($z$\,=\,3). Our KMOS galaxies are off the $z\sim0$ TF relation by $\sim2.6$\,$\sigma$, but are very well fitted by the $z\sim1-2$ TF relation. The clear group members (all within a 3\,Mpc diameter) seem to have slightly higher masses for a fixed velocity, but the two samples differ by only 1\,$\sigma$, and thus this is likely driven by the low number statistics and the higher masses of the group members.}
\label{fig:TF}
\end{figure}

\section{Results and Discussion}
\label{sec:disc}

From the full target sample, 13 are at $z$\,=\,0.813 and within 1000\,km\,s$^{-1}$ of each other, thus identifying the redshift of this group of star-forming galaxies. As a comparison, the FWHM of the NB filter recovers H$\alpha$ emitters within $\sim3000$\,km\,s$^{-1}$. Moreover, 7 galaxies are found within a 3\,Mpc diameter. All of these group members show higher masses and lower sSFRs than the rest of the sample (see Fig.1).

Turning to the line ratios, we use the galaxy integrated [N{\sc ii}]/H$\alpha$ emission line ratio to infer the metallicity of the gas. Across the full sample, the average ratio is [N{\sc ii}]\,/\,H$\alpha$\,=\,$0.32\pm0.13$, consistent with the sample of $\sim $\,100 HiZELS galaxies at a similar redshift \citep{Stott13b}. The [N{\sc ii}]/H$\alpha$ line ratio can be used to determine the metallicity of our galaxies (Oxygen abundance), [12\,+\,log(O/H)], by using the conversion of Pettini \& Pagel (2004), appropriate for high redshift galaxies: 12\,+\,log(O/H)\,=\,8.9\,+\,0.57\,log([N{\sc ii}]\,/\,H$\alpha$). The galaxies in our sample have a median metallicity of $8.62\pm0.07$, which is slightly lower than solar, but still consistent with the solar value of 8.66$\pm$0.05. Our KMOS galaxies have metallicities consistent with those in Swinbank et al (2012a), who derive 12\,+\,log(O/H)\,=\,$8.58\pm0.07$ for H$\alpha$-selected samples at $z\sim $\,0.84--1.47. Our KMOS galaxies are also very well-fitted by the Mass-SFR-Metallicity fundamental-plane for $z\sim1$ galaxies derived by \cite{Stott13b}; this means that our galaxies do not show any significant difference from those generally found in the field. We note that group members are slightly more metal-rich than the galaxies in the outskirts and/or field, but we find this is solely driven by such sources also being more massive (see Fig. 1). At a fixed mass, there is no difference in metallicities and we find no environmental effect in the mass-metallicity relation between these group galaxies and those in the field.

Of the sixteen galaxies in our sample, thirteen are classified as disks, whilst the remainder do not have regular dynamics (either unresolved or merging systems). This corresponds to a fraction of disks of 75\,$\pm$\,8\%, which is in excellent agreement with \cite{Sobral09} who found that the rest-frame $R$-band morphologies (measured from \emph{HST}) of $\sim $\,80\% of $z$\,=\,0.84 H$\alpha$ selected star-forming galaxies are disk-like. It is also consistent with the results from \cite{Stott13a}, who used $H$-band data to derive the Sersic profile of hundreds of H$\alpha$ selected galaxies at $z=0.4-2.23$, including $z=0.84$. The fraction of rotating systems within our sample is also consistent with that found from other H$\alpha$ IFU surveys of high-redshift star-forming galaxies in the field \citep[e.g.\ ][]{ForsterSchreiber09,Jones10,Wisnioski11,Swinbank12a}. Our results confirm that the majority of the ``representative'' star-forming galaxies at $z\sim1$ are disks and add to the picture that it is the evolution of disks that is responsible for the decline of the star formation rate density at least since $z\sim1$. Interestingly, among the three sources which are not well-fitted by rotating disk models, two are likely at the (opposite) edges of this structure. All sources within the 3\,Mpc diameter are disk-like.

We use the inclination-corrected rotation speeds and stellar masses of the galaxies in our sample to investigate the Tully-Fisher (TF) relation for our $z\sim0.8$ galaxies and show our results in Fig.~\ref{fig:TF}. The stellar masses and velocities from the literature in Fig.~\ref{fig:TF} on this plot have been estimated in a fully consistent way, and these values (or corrections, where necessary) are presented in \cite{Swinbank12a}. We also show the TF relation fits at $z=0$ and $z=1-2$ for reference/comparison; these have been derived from the compilation of star-forming galaxies in \cite{Swinbank12a}. Due to our relatively small sample we do not attempt to fit a relation to our data, but the $z=1-2$ fit derived in \cite{Swinbank12a} provides a much better fit to our data than the $z=0$ TF relation. In fact, as Fig.~\ref{fig:TF} shows, the $z\sim0.8$ KMOS sources in our sample have, on average, slightly lower stellar masses for a given velocity when compared to local galaxies, in agreement with previous study \citep[e.g.][]{Swinbank06a,Yang08,Cresci09,Puech10,Swinbank12a}, \citep[but see also][]{Jones10,Miller11,Miller12}. In order to quantify the statistical significance of this offset from the $z=0$ TF relation, we take the full $z=0$ sample, randomly select 10 galaxies, fix the slope of the TF relation at the $z=0$ value, and fit the normalization using the sub-sample. We repeat this process 10000 times, and then do the same for our KMOS sample. We find that the normalization of the two differs by about 2.6$\sigma$. By applying the same procedure to the $z\sim1$ and $z\sim2$ samples, we find them to be indistinguishable from our KMOS sample and thus fully consistent with being drawn from the same larger sample. By separating our galaxies between those confirmed to reside in the $z=0.813$ group and those outside the group, we find that group galaxies may be slightly more massive at a fixed velocity, when compared to field galaxies, but this is only a 1$\sigma$ effect and thus likely driven by a combination of group galaxies being more massive (irrespectively of their velocities) and low number statistics. Therefore, field and group galaxies present the same TF relation.

\section{Conclusions}

We presented the spatially resolved H$\alpha$ dynamics of sixteen star-forming galaxies at $z\sim0.81$ using the new KMOS multi-object integral field spectrograph on the ESO VLT. We confirm and identify a rich group of star-forming galaxies at $z= $\,0.813\,$\pm$\,0.003, with thirteen galaxies within 1000\,km\,s$^{-1}$ of each other, and 7 within a diameter of 3\,Mpc. Overall, our $\sim$\,SFR$^*$ (typical) KMOS star-forming galaxies span a range of specific star formation rate of sSFR\,=\,0.2--1.1\,Gyr$^{-1}$ and have a median metallicity very close to solar of 12\,+\,log(O/H)\,=\,8.62\,$\pm$\,0.06. We measure the spatially resolved H$\alpha$ dynamics of the galaxies in our sample and show that thirteen out of sixteen galaxies can be described by rotating disks and use the data to derive inclination corrected rotation speeds of 50--275\,km\,s$^{-1}$. The fraction of disks within our sample is 75\,$\pm$\,8\%, consistent with previous results based on \emph{HST} morphologies of H$\alpha$ selected galaxies at $z\sim $\,1 and confirming that disks dominate the star formation rate density at $z\sim $\,1. Our KMOS galaxies are very well-fitted by the field Mass-SFR-Metallicity relation at $z\sim1$ \citep{Stott13b}. Galaxies in the group have slightly higher metallicities, but also higher masses, and thus still completely consistent with the Mass-SFR-Metallicity relation at $z\sim1$. We find that our $z\sim0.81$ KMOS galaxies are off the $z=0$ TF relation by 2.6\,$\sigma$, but that they are very well-fitted by the $z\sim1-2$ TF relation, with our sample being statistically indistinguishable from other $z\sim1-2$ samples. We conclude that while many of our KMOS galaxies reside in a relatively dense region/group environment, they have, nevertheless, similar properties to galaxies residing in typical/field densities. Thus, apart from having, on average, higher stellar masses and lower sSFRs, our group galaxies at $z=0.81$ present the same mass-metallicity and TF relation as $z\sim1-2$ field galaxies, and are all disk galaxies.

\section*{acknowledgments}

We thank the referee for many helpful comments and suggestions which greatly improved the clarity and quality of this work. DS acknowledges financial support from the Netherlands Organisation for Scientific research (NWO) through a Veni fellowship and also funding from the European Communityâ Seventh Framework Programme (FP7/2007-2013) under grant agreement number RG226604 (OPTICON) which allowed access to CFHT time (proposals: 11BO29 \& 12AO19). AMS gratefully acknowledges an STFC Advanced Fellowship through grant number ST/H005234/1. IRS, JPS and RGB acknowledge support from the UK Science and Technology Facilities Council (STFC) under ST/I001573/1. IRS acknowledges STFC (ST/J001422/1), the ERC Advanced Investigator programme DUSTYGAL and a Royal Society/Wolfson Merit Award. PNB acknowledges support from STFC. RMS acknowledges support from the grant ST/1001573/1. The data presented here are based on observations with the KMOS spectrograph on the ESO/VLT under program 60.A-9460 and can be accessed through the ESO data archive. The authors also wish to acknowledge the help from Michael Hilker in preparing the KMOS observations.

\bibliographystyle{apj}

\begin{thebibliography}{47}
\expandafter\ifx\csname natexlab\endcsname\relax\def\natexlab#1{#1}\fi

\bibitem[{{Bower} {et~al.}(2012){Bower}, {Benson}, \& {Crain}}]{Bower12}
{Bower}, R.~G., {Benson}, A.~J., \& {Crain}, R.~A. 2012, \mnras, 422, 2816

\bibitem[{{Charbonneau}(1995)}]{Charbonneau95}
{Charbonneau}, P. 1995, \apjs, 101, 309

\bibitem[{{Cirasuolo} {et~al.}(2010){Cirasuolo}, {McLure}, {Dunlop}, {Almaini},
  {Foucaud}, \& {Simpson}}]{Cirasuolo10}
{Cirasuolo}, M., {McLure}, R.~J., {Dunlop}, J.~S., {Almaini}, O., {Foucaud},
  S., \& {Simpson}, C. 2010, \mnras, 401, 1166

\bibitem[{{Conselice} {et~al.}(2009){Conselice}, {Yang}, \&
  {Bluck}}]{Conselice09}
{Conselice}, C.~J., {Yang}, C., \& {Bluck}, A.~F.~L. 2009, \mnras, 394, 1956

\bibitem[{{Courteau}(1997)}]{Courteau97}
{Courteau}, S. 1997, \aj, 114, 2402

\bibitem[{{Cresci} {et~al.}(2009){Cresci}, {Hicks}, {Genzel}, {Schreiber},
  {Davies}, {Bouch{\'e}}, {Buschkamp}, \& {Genel, S. et al}}]{Cresci09}
{Cresci}, G., {Hicks}, E.~K.~S., {Genzel}, R., {Schreiber}, N.~M.~F., {Davies},
  R., {Bouch{\'e}}, N., {Buschkamp}, P., {Genel, S. et al}. 2009, \apj, 697,
  115

\bibitem[{{Davies}(2007)}]{Davies07}
{Davies}, R.~I. 2007, \mnras, 375, 1099

\bibitem[{{Davies} {et~al.}(2013){Davies}, {Agudo Berbel}, {Wiezorrek},
  {Cirasuolo}, {Forster Schreiber}, {Jung}, {Muschielok}, {Ott}, {Ramsay},
  {Schlichter}, {Sharples}, \& {Wegner}}]{Davies13}
{Davies}, R.~I., {Agudo Berbel}, A., {Wiezorrek}, E., {Cirasuolo}, M., {Forster
  Schreiber}, N.~M., {Jung}, Y., {Muschielok}, B., {Ott}, T., {Ramsay}, S.,
  {Schlichter}, J., {Sharples}, R., \& {Wegner}, M. 2013, \aap, 558, 17

\bibitem[{{Dekel} {et~al.}(2009){Dekel}, {Birnboim}, {Engel}, {Freundlich},
  {Goerdt}, {Mumcuoglu}, {Neistein}, \& {Pichon, C et al.}}]{Dekel09}
{Dekel}, A., {Birnboim}, Y., {Engel}, G., {Freundlich}, J., {Goerdt}, T.,
  {Mumcuoglu}, M., {Neistein}, A., {Pichon, C et al.} 2009, \nat, 457, 451

\bibitem[{{Dom{\'{\i}}nguez} {et~al.}(2013){Dom{\'{\i}}nguez}, {Siana},
  {Henry}, {Scarlata}, {Bedregal}, {Malkan}, {Atek}, {Ross}, {Colbert},
  {Teplitz}, {Rafelski}, {McCarthy}, {Bunker}, {Hathi}, {Dressler}, {Martin},
  \& {Masters}}]{Dominguez13}
{Dom{\'{\i}}nguez}, A., {Siana}, B., {Henry}, A.~L., {Scarlata}, C.,
  {Bedregal}, A.~G., {Malkan}, M., {Atek}, H., {Ross}, N.~R., {Colbert}, J.~W.,
  {Teplitz}, H.~I., {Rafelski}, M., {McCarthy}, P., {Bunker}, A., {Hathi},
  N.~P., {Dressler}, A., {Martin}, C.~L., \& {Masters}, D. 2013, \apj, 763, 145
  
\bibitem[{{Dutton} {et~al.}(2011){Dutton}, {van den Bosch}, {Faber}, {Simard},
  {Kassin}, {Koo}, {Bundy}, \& {Huang, J. et al.}}]{Dutton11}
{Dutton}, A.~A., {van den Bosch}, F.~C., {Faber}, S.~M., {Simard}, L.,
  {Kassin}, S.~A., {Koo}, D.~C., {Bundy}, K., \& {Huang, J. et al.} 2011,
  \mnras, 410, 1660

\bibitem[{{Elmegreen} {et~al.}(2009){Elmegreen}, {Elmegreen}, {Fernandez}, \&
  {Lemonias}}]{Elmegreen09}
{Elmegreen}, B.~G., {Elmegreen}, D.~M., {Fernandez}, M.~X., \& {Lemonias},
  J.~J. 2009, \apj, 692, 12

\bibitem[{{F{\"o}rster Schreiber} {et~al.}(2009{\natexlab{a}}){F{\"o}rster
  Schreiber}, {Genzel}, {Bouch{\'e}}, {Cresci}, {Davies}, {Buschkamp},
  {Shapiro}, \& {et al.}}]{Schreiber09}
{F{\"o}rster Schreiber}, N.~M., {Genzel}, R., {Bouch{\'e}}, N., {Cresci}, G.,
  {Davies}, R., {Buschkamp}, P., {Shapiro}, K., {et al.} 2009{\natexlab{a}},
  \apj, 706, 1364

\bibitem[{{F{\"o}rster Schreiber} {et~al.}(2009{\natexlab{b}}){F{\"o}rster
  Schreiber}, {Genzel}, {Bouch{\'e}}, {Cresci}, {Davies}, {Buschkamp},
  {Shapiro}, \& {Tacconi, L.~J.. et al.}}]{ForsterSchreiber09}
{F{\"o}rster Schreiber}, N.~M., {Genzel}, R., {Bouch{\'e}}, N., {Cresci}, G.,
  {Davies}, R., {Buschkamp}, P., {Shapiro}, K., {Tacconi, L.~J.. et al.}
  2009{\natexlab{b}}, \apj, 706, 1364

\bibitem[{{Garn} \& {Best}(2010)}]{GarnBest10}
{Garn}, T. \& {Best}, P.~N. 2010, \mnras, 409, 421

\bibitem[{{Garn} {et~al.}(2010){Garn}, {Sobral}, {Best}, {Geach}, {Smail},
  {Cirasuolo}, {Dalton}, {Dunlop}, {McLure}, \& {Farrah}}]{Garn10}
{Garn}, T., {Sobral}, D., {Best}, P.~N., {Geach}, J.~E., {Smail}, I.,
  {Cirasuolo}, M., {Dalton}, G.~B., {Dunlop}, J.~S., {McLure}, R.~J., \&
  {Farrah}, D. 2010, \mnras, 402, 2017

\bibitem[{{Geach} {et~al.}(2008){Geach}, {Smail}, {Best}, {Kurk}, {Casali},
  {Ivison}, \& {Coppin}}]{Geach08}
{Geach}, J.~E., {Smail}, I., {Best}, P.~N., {Kurk}, J., {Casali}, M., {Ivison},
  R.~J., \& {Coppin}, K. 2008, \mnras, 388, 1473

\bibitem[{{Geach} {et~al.}(2012){Geach}, {Sobral}, {Hickox}, {Wake}, {Smail},
  {Best}, {Baugh}, \& {Stott}}]{Geach12}
{Geach}, J.~E., {Sobral}, D., {Hickox}, R.~C., {Wake}, D.~A., {Smail}, I.,
  {Best}, P.~N., {Baugh}, C.~M., \& {Stott}, J.~P. 2012, \mnras, 426, 679

\bibitem[{{Genzel} {et~al.}(2010){Genzel}, {Tacconi}, {Gracia-Carpio},
  {Sternberg}, {Cooper}, {Shapiro}, {Bolatto}, {Bouche, N. et
  al.,}}]{Genzel10}
{Genzel}, R., {Tacconi}, L.~J., {Gracia-Carpio}, J., {Sternberg}, A., {Cooper},
  M.~C., {Shapiro}, K., {Bolatto}, A., {Bouche, N. et al.,}. 2010, \mnras,
  407, 2091

\bibitem[{{Gilbank} {et~al.}(2011){Gilbank}, {Bower}, {Glazebrook}, {Balogh},
  {Baldry}, {Davies}, {Hau}, {Li}, {McCarthy}, \& {Sawicki}}]{Gilbank11}
{Gilbank}, D.~G., {Bower}, R.~G., {Glazebrook}, K., {Balogh}, M.~L., {Baldry},
  I.~K., {Davies}, G.~T., {Hau}, G.~K.~T., {Li}, I.~H., {McCarthy}, P., \&
  {Sawicki}, M. 2011, \mnras, 414, 304

\bibitem[{{Glazebrook}(2013)}]{Glazebrook13}
{Glazebrook}, K. 2013, arXiv:1305.2469
  
  \bibitem[{{Gnerucci} {et~al.}(2011){Gnerucci}, {Marconi}, {Cresci}, {Maiolino},
  {Mannucci}, {Calura}, {Cimatti}, \& {Cocchia, F. et al.}}]{Gnerucci11}
{Gnerucci}, A., {Marconi}, A., {Cresci}, G., {Maiolino}, R., {Mannucci}, F.,
  {Calura}, F., {Cimatti}, A., \& {Cocchia, F. et al.} 2011, \aap, 528, A88
  
  
\bibitem[{{Gon{\c c}alves} {et~al.}(2010){Gon{\c c}alves}, {Basu-Zych}, {Overzier}, {Martin},
  {Law}, {Schiminovich}, {Wyder}, \& {Mallery et al.}}]{Goncalves10}
{Gon{\c c}alves}, T.~S.,  {Basu-Zych}, A., {Overzier}, R., {Martin}, D.~C.,
  {Law}, D.~R., {Schiminovich}, D., {Wyder}, R.~K., \& {Mallery et al.} 2010, \apj, 724, 1373
  
  
\bibitem[{{Hopkins}(2012)}]{Hopkins12b}
{Hopkins}, P.~F. 2012, \mnras, 423, 2016

\bibitem[{{Ibar} {et~al.}(2013){Ibar}, {Sobral}, {Best}, {Ivison}, {Smail},
  {Arumugam}, {Berta}, {B{\'e}thermin}, {Bock}, \& {et al.}}]{Ibar13}
{Ibar}, E., {Sobral}, D., {Best}, P.~N., {Ivison}, R.~J., {Smail}, I.,
  {Arumugam}, V., {Berta}, S., {B{\'e}thermin}, M., {Bock}, J., {et al.}
  2013, \mnras, 434, 3218

\bibitem[{{Jones} {et~al.}(2010){Jones}, {Swinbank}, {Ellis}, {Richard}, \&
  {Stark}}]{Jones10}
{Jones}, T.~A., {Swinbank}, A.~M., {Ellis}, R.~S., {Richard}, J., \& {Stark},
  D.~P. 2010, \mnras, 404, 1247

\bibitem[{{Karim} {et~al.}(2011){Karim}, {Schinnerer},
  {Mart{\'{\i}}nez-Sansigre}, {Sargent}, {van der Wel}, {Rix}, {Ilbert},
  {Smol{\v c}i{\'c}}, {Carilli}, {Pannella}, {Koekemoer}, {Bell}, \&
  {Salvato}}]{Karim11}
{Karim}, A., {Schinnerer}, E., {Mart{\'{\i}}nez-Sansigre}, A., {Sargent},
  M.~T., {van der Wel}, A., {Rix}, H.-W., {Ilbert}, O., {Smol{\v c}i{\'c}}, V.,
  {Carilli}, C., {Pannella}, M., {Koekemoer}, A.~M., {Bell}, E.~F., \&
  {Salvato}, M. 2011, \apj, 730, 61

\bibitem[{{Krajnovi{\'c}} {et~al.}(2006){Krajnovi{\'c}}, {Cappellari}, {de
  Zeeuw}, \& {Copin}}]{Krajnovic07}
{Krajnovi{\'c}}, D., {Cappellari}, M., {de Zeeuw}, P.~T., \& {Copin}, Y. 2006,
  \mnras, 366, 787

\bibitem[{{Livermore} {et~al.}(2012){Livermore}, {Jones}, {Richard}, {Bower},
  {Ellis}, {Swinbank}, {Rigby}, \& {Smail, I. et al.}}]{Livermore12a}
{Livermore}, R.~C., {Jones}, T., {Richard}, J., {Bower}, R.~G., {Ellis}, R.~S.,
  {Swinbank}, A.~M., {Rigby}, J.~R., {Smail, I. et al.} 2012, \mnras, 427,
  688

\bibitem[{{Madau} {et~al.}(1996){Madau}, {Ferguson}, {Dickinson}, {Giavalisco},
  {Steidel}, \& {Fruchter}}]{Madau96}
{Madau}, P., {Ferguson}, H.~C., {Dickinson}, M.~E., {Giavalisco}, M.,
  {Steidel}, C.~C., \& {Fruchter}, A. 1996, \mnras, 283, 1388

\bibitem[{{Miller} {et~al.}(2011){Miller}, {Bundy}, {Sullivan}, {Ellis}, \&
  {Treu}}]{Miller11}
{Miller}, S.~H., {Bundy}, K., {Sullivan}, M., {Ellis}, R.~S., \& {Treu}, T.
  2011, \apj, 741, 115

\bibitem[{{Miller} {et~al.}(2012){Miller}, {Ellis}, {Sullivan}, {Bundy},
  {Newman}, \& {Treu}}]{Miller12}
{Miller}, S.~H., {Ellis}, R.~S., {Sullivan}, M., {Bundy}, K., {Newman}, A.~B.,
  \& {Treu}, T. 2012, \apj, 753, 74

\bibitem[{{Pizagno} {et~al.}(2005){Pizagno}, {Prada}, {Weinberg}, {Rix},
  {Harbeck}, {Grebel}, {Bell}, \& {Brinkmann, J. et al.}}]{Pizagno05}
{Pizagno}, J., {Prada}, F., {Weinberg}, D.~H., {Rix}, H.-W., {Harbeck}, D.,
  {Grebel}, E.~K., {Bell}, E.~F., {Brinkmann, J. et al.} 2005, \apj, 633,
  844
  
  \bibitem[{{Puech} {et~al.}(2010){Puech}, {Hammer}, {Flores et al.}}]{Puech10}
{Puech}, J., {Hammer}, F., {Flores H. et al.} 2010, \aap, 510, A68

\bibitem[{{Puget} {et~al.}(2004){Puget}, {Stadler}, {Doyon}, {et al.},}]{Puget2004}
{Puget}, P., {Stadler}, E., {Doyon}, R., {et al.} 2004, in Society of Photo-Optical
  Instrumentation Engineers (SPIE) Conference Series, Vol. 5492, Society of
  Photo-Optical Instrumentation Engineers (SPIE) Conference Series, ed.
  A.~F.~M. {Moorwood} \& M.~{Iye}, 978--987
  
\bibitem[{{Rodighiero} {et~al.}(2011){Rodighiero}, {Daddi}, {Baronchelli},
  {Cimatti}, {Renzini}, {Aussel}, {Popesso}, \& {Lutz, D. et
  al.}}]{Rodighiero11}
{Rodighiero}, G., {Daddi}, E., {Baronchelli}, I., {Cimatti}, A., {Renzini}, A.,
  {Aussel}, H., {Popesso}, P., {Lutz, D. et al.} 2011, \apjl, 739, L40

\bibitem[{{Shapiro} {et~al.}(2008){Shapiro}, {Genzel}, {F{\"o}rster Schreiber},
  {Tacconi}, {Bouch{\'e}}, {Cresci}, {Davies}, \& {Eisenhauer, F. et
  al.}}]{Shapiro08}
{Shapiro}, K.~L., {Genzel}, R., {F{\"o}rster Schreiber}, N.~M., {Tacconi},
  L.~J., {Bouch{\'e}}, N., {Cresci}, G., {Davies}, R., {Eisenhauer, F. et
  al.} 2008, \apj, 682, 231

\bibitem[{{Sharples} {et~al.}(2013){Sharples}, {Bender}, {Agudo Berbel},
  {Bezawada}, {Castillo}, {Cirasuolo}, {Davidson}, {Davies}, {et
  al.}}]{Sharples13}
{Sharples}, R., {Bender}, R., {Agudo Berbel}, A., {Bezawada}, N., {Castillo},
  R., {Cirasuolo}, M., {Davidson}, G., {Davies}, R., {et al.} 2013, The
  Messenger, 151, 21

\bibitem[{{Sobral} {et~al.}(2010){Sobral}, {Best}, {Geach}, {Smail},
  {Cirasuolo}, {Garn}, {Dalton}, \& {Kurk}}]{Sobral10}
{Sobral}, D., {Best}, P.~N., {Geach}, J.~E., {Smail}, I., {Cirasuolo}, M.,
  {Garn}, T., {Dalton}, G.~B., \& {Kurk}, J. 2010, \mnras, 404, 1551

\bibitem[{{Sobral} {et~al.}(2009){Sobral}, {Best}, {Geach}, {Smail}, {Kurk},
  {Cirasuolo}, {Casali}, \& {Ivison, R.~J. et al.}}]{Sobral09}
{Sobral}, D., {Best}, P.~N., {Geach}, J.~E., {Smail}, I., {Kurk}, J.,
  {Cirasuolo}, M., {Casali}, M., {Ivison, R.~J. et al.} 2009, \mnras, 398,
  75

\bibitem[{{Sobral} {et~al.}(2012){Sobral}, {Best}, {Matsuda}, {Smail}, {Geach},
  \& {Cirasuolo}}]{Sobral12}
{Sobral}, D., {Best}, P.~N., {Matsuda}, Y., {Smail}, I., {Geach}, J.~E., \&
  {Cirasuolo}, M. 2012, \mnras, 420, 1926

\bibitem[{{Sobral} {et~al.}(2011){Sobral}, {Best}, {Smail}, {Geach},
  {Cirasuolo}, {Garn}, \& {Dalton}}]{Sobral11}
{Sobral}, D., {Best}, P.~N., {Smail}, I., {Geach}, J.~E., {Cirasuolo}, M.,
  {Garn}, T., \& {Dalton}, G.~B. 2011, \mnras, 411, 675
  
  \bibitem[{{Sobral} {et~al.}(2013{\natexlab{a}}){Sobral}, {Smail}, {Best},
  {Geach}, {Matsuda}, {Stott}, {Cirasuolo}, \& {Kurk}}]{Sobral13}
{Sobral}, D., {Smail}, I., {Best}, P.~N., {Geach}, J.~E., {Matsuda}, Y.,
  {Stott}, J.~P., {Cirasuolo}, M., \& {Kurk}, J. 2013{\natexlab{a}}, \mnras,
  428, 1128

\bibitem[{{Sobral} {et~al.}(2013{\natexlab{b}}){Sobral}, {Best}, {Smail},
  {Mobasher}, {Stott}, \& {Nisbet}}]{Sobral13b}
{Sobral}, D., {Best}, P.~N., {Smail}, I., {Mobasher}, B., {Stott}, J., \&
  {Nisbet}, D. 2013{\natexlab{b}}, \mnras, in press [arXiv:1311.1503]

\bibitem[{{Sobral} {et~al.}(2013{\natexlab{c}}){Sobral}, {Matthee}, {Kim},
  {et al.}}]{Sobral13c}
{Sobral}, D., {Matthee}, J., {Kim}, I., {et al.} 2013{\natexlab{c}}, \mnras,
  in prep.

\bibitem[{{Stott} {et~al.}(2013{\natexlab{a}}){Stott}, {Sobral}, {Smail}, {Bower}, {Best}, \&
  {Geach}}]{Stott13a}
{Stott}, J.~P., {Sobral}, D., {Smail}, I., {Bower}, R., {Best}, P.~N., \&
  {Geach}, J.~E. 2013{\natexlab{a}}, \mnras, 430, 1158
  
  \bibitem[{{Stott} {et~al.}(2013{\natexlab{b}}){Stott}, {Sobral}, {Bower}, {Smail}, {Best},
  {et al.}}]{Stott13b}
{Stott}, J.~P., {Sobral}, D., {Bower}, R., {Smail}, I., {Best}, P.~N.,
  {et al.} 2013{\natexlab{b}}, \mnras, 436, 1130

\bibitem[{{Swinbank} {et~al.}(2006){Swinbank}, {Bower}, {Smith}, {Smail},
  {Kneib}, {Ellis}, {Stark}, \& {Bunker}}]{Swinbank06a}
{Swinbank}, A.~M., {Bower}, R.~G., {Smith}, G.~P., {Smail}, I., {Kneib}, J.-P.,
  {Ellis}, R.~S., {Stark}, D.~P., \& {Bunker}, A.~J. 2006, \mnras, 368, 1631

\bibitem[{{Swinbank} {et~al.}(2012{\natexlab{a}}){Swinbank}, {Sobral}, {Smail},
  {Geach}, {Best}, {McCarthy}, {Crain}, \& {Theuns}}]{Swinbank12a}
{Swinbank}, A.~M., {Sobral}, D., {Smail}, I., {Geach}, J.~E., {Best}, P.~N.,
  {McCarthy}, I.~G., {Crain}, R.~A., \& {Theuns}, T. 2012{\natexlab{a}},
  \mnras, 426, 935

\bibitem[{{Swinbank} {et~al.}(2012{\natexlab{b}}){Swinbank}, {Smail}, {Sobral},
  {Theuns}, {Best}, \& {Geach}}]{Swinbank12b}
{Swinbank}, M., {Smail}, I., {Sobral}, D., {Theuns}, T., {Best}, P., \&
  {Geach}, J. 2012{\natexlab{b}}, \apj, 760, 130

\bibitem[{{Wisnioski} {et~al.}(2011){Wisnioski}, {Glazebrook}, {Blake},
  {Wyder}, {Martin}, {Poole}, {Sharp}, \& {Couch, W. et al.}}]{Wisnioski11}
{Wisnioski}, E., {Glazebrook}, K., {Blake}, C., {Wyder}, T., {Martin}, C.,
  {Poole}, G.~B., {Sharp}, R.,  {Couch, W. et al.} 2011, \mnras, 417, 2601
  
\bibitem[{{Yang} {et~al.}(2008){Yang}, {Flores}, {Hammer et al.}}]{Yang08}
{Yang}, E., {Flores}, H., {Blake}, C., {Hammer, F. et al.}, 2008, \aap, 2008,  477, 789
  

\end{thebibliography}

\end{document}